\documentclass{PoS}

\usepackage{graphicx}
\usepackage{dcolumn}
\usepackage{bm}

\title{Finite $ma$ Errors of the Overlap Fermion}

\ShortTitle{Overlap $ma$ Errors}

\author{\speaker{S.J. Dong} and K.F. Liu\\
        Dept. of Physics and Astronomy, University of Kentucky, Lexington, KY 40506, USA\\
        E-mail: \email{super124@uky.edu}}

\abstract{We shall discuss the finite $ma$ errors of the overlap
fermion in this talk. We present results on the speed of light from the
dispersion relation and the hyperfine splitting between the vector
and pseudoscalar mesons as a function to $ma$ to reveal the
$m\Lambda_{QCD}a^2$ and $m^2a^2$ errors. We conclude from this
study that one should be limited to using $ma$ less than 0.5 in
order to keep the systematic $ma$ errors below a few percent level.}

\FullConference{The XXV International Symposium on Lattice Field Theory\\
		 July 30-4 August 2007\\
		 Regensburg, Germany}

\begin{document}
In the last few years, there have been a number of studies to 
check how well current algebra relations are satisfied with various 
numerical approximations for the overlap fermion and how feasible it is to 
carry out large scale calculations with realistically small quark masses. 
Although it takes two orders of magnitude more time to 
compute the propagators than those of the Wilson-type fermions, the
overlap fermion has a host of desirable features, such as the fact that 
there are no $O(a)$ errors, no additive renormalization for the quark masses, 
no mixing between different chiral sectors, and that the chiral Ward 
identity and other current algebra relations make the non-perturbative 
renormalization easier, etc. In addition, there are a number of
pleasant surprises in that the $O(a^2)$ errors, as judged from
hadron masses, is small~\cite{dll00} and is about the smallest among
the fermion actions studied in the quenched simulations~\cite{dmz05}. 
The overlap fermion is local for lattice spacing as coarse as 0.2 fm 
with the range being roughly one lattice spacing for the Euclidean 
distance~\cite{dmz05}.
The renormalization factors from the chiral Ward identity and the
regularization independent scheme~\cite{mps95} have very little
dependence on $ma$~\cite{zmd05} for $ma$ as large as 0.7. The
speed of light as calculated from the dispersion relation deviates
from unity appreciably only for $ma$ larger than $\sim 0.55$
~\cite{ld05}. 

\begin{figure}[ht]
\vspace*{8.5cm} 
\includegraphics{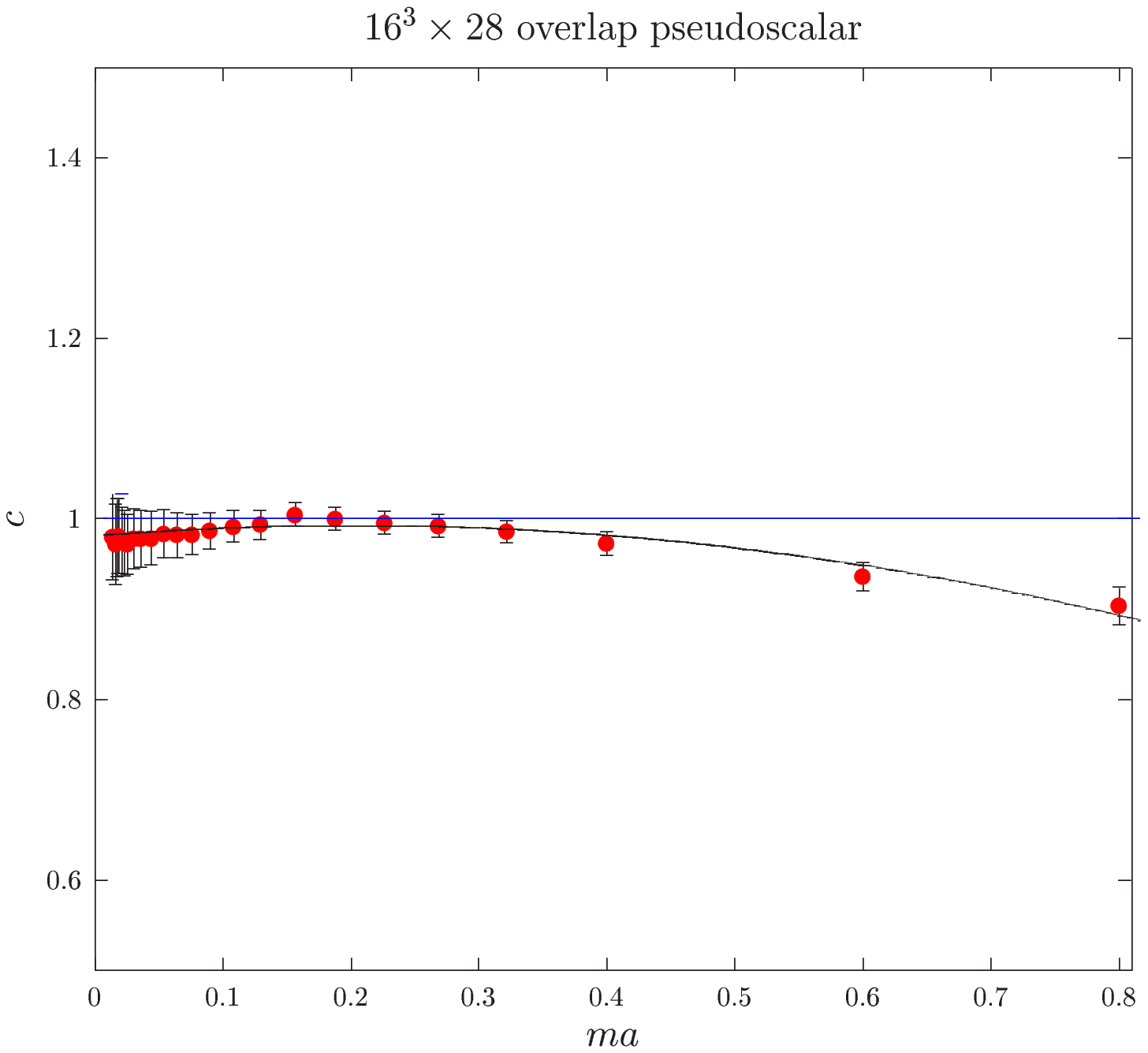} 
\vspace*{8.5cm} 
\includegraphics{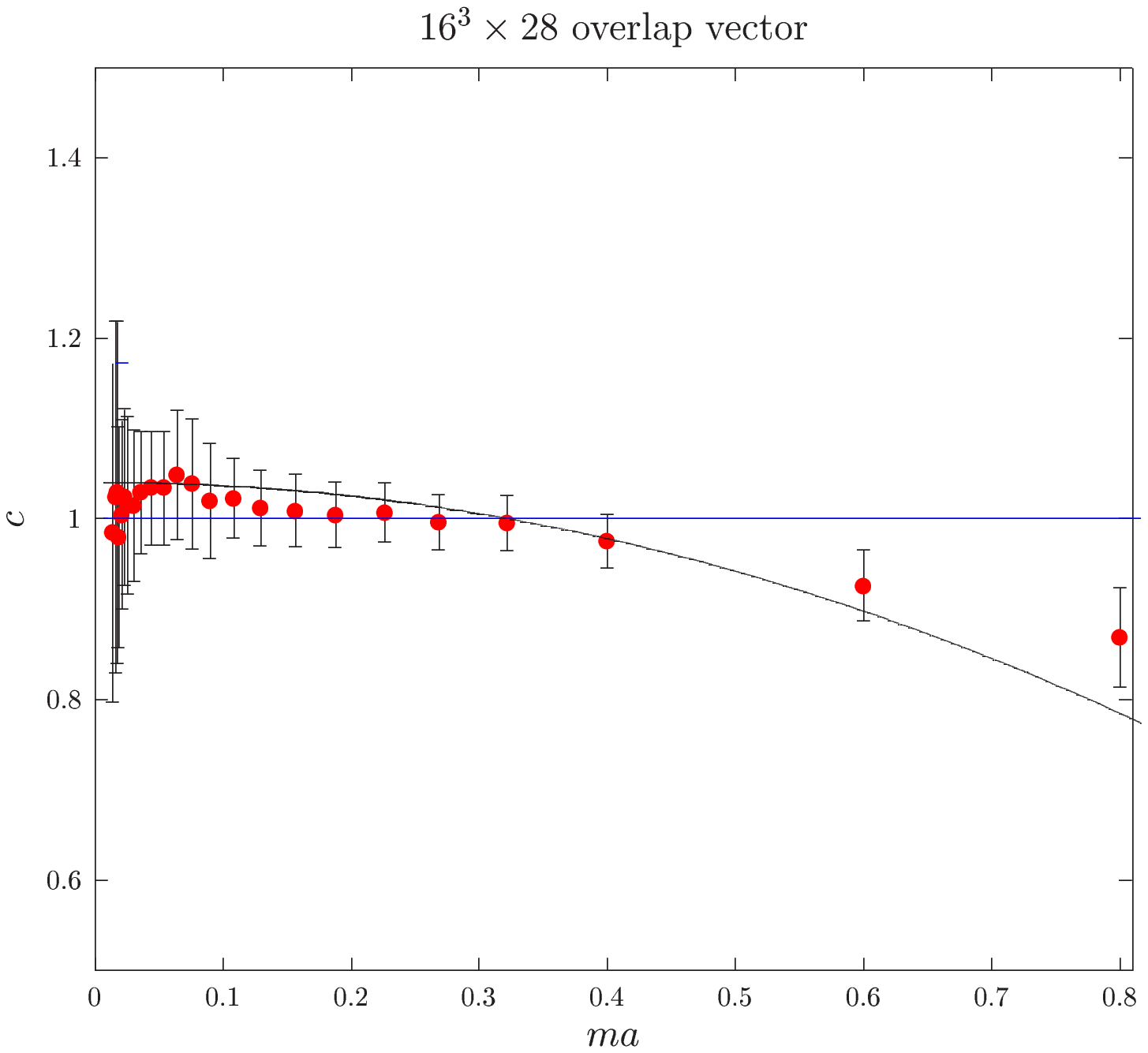}
\caption{The effective
speed of light $c$ from the pseudoscalar-meson (upper panel) and
vector-meson (lower panel) dispersion relations
as a function of $ma$.} \label{disp_pi}
\end{figure}

   It was emphasized that the effective quark propagator for the
overlap fermion has the same form as that in the continuum, i.e.
the inverse effective propagator is just an anti-hermitian Dirac operator
plus the bare quark mass term~\cite{ld05}. As such, the overlap
fermion is equally applicable to the heavy as well as the light
quarks. The only practical concern is how large the $ma$ errors
are for the heavy quarks. Thus, it is essential to assess the $ma$
errors before one can confidently apply the overlap formalism to
heavy quarks for a specific $ma$. For this purpose, we present
results on the dispersion relation and the hyperfine splitting
between the vector and pseudoscalar mesons as a function of $ma$
to reveal the $m\Lambda_{QCD}a^2$ and $m^2a^2$ errors.

\begin{figure}[h]
\vspace*{8cm} \includegraphics{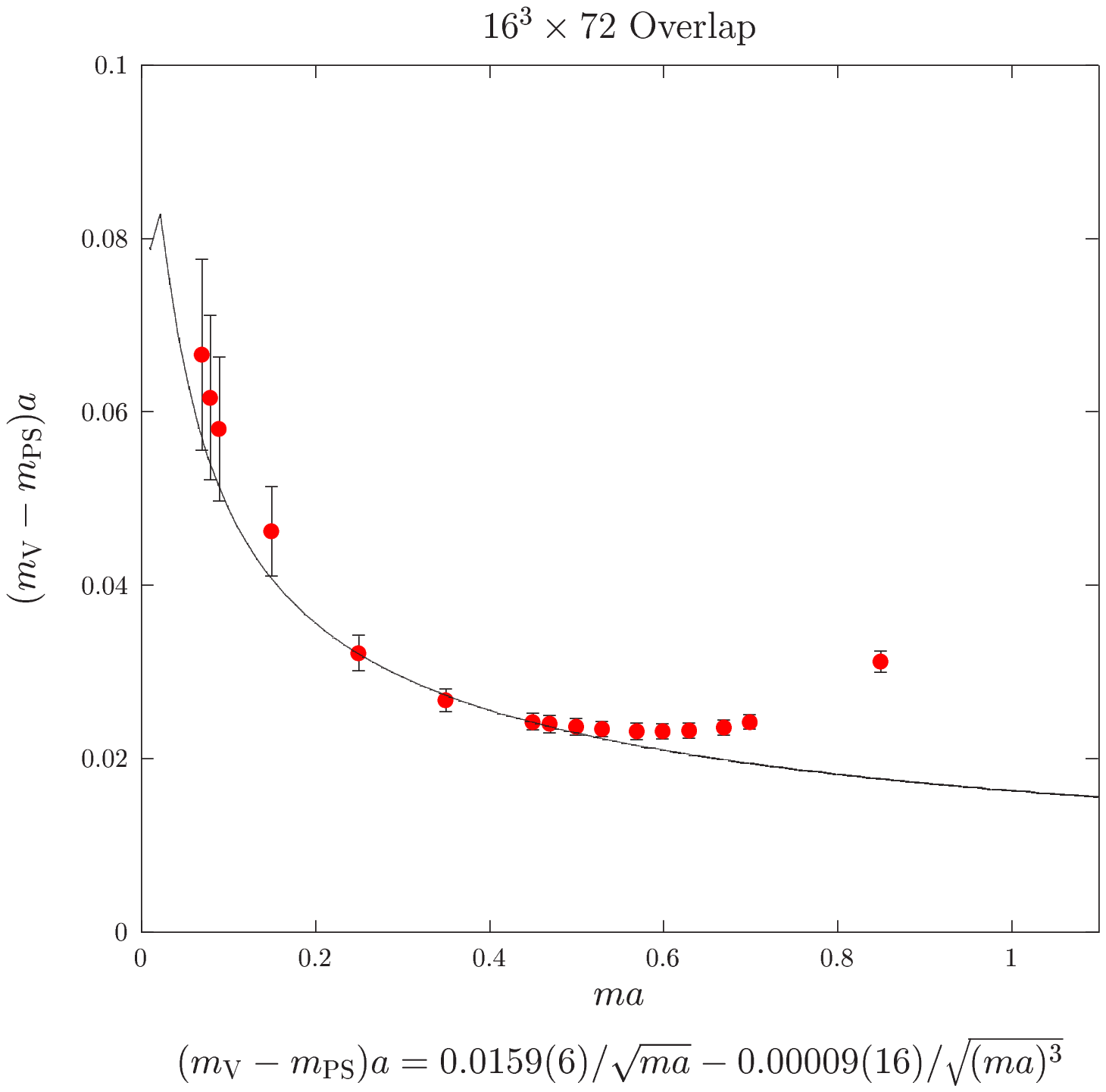} \vspace*{0.6cm}
\vspace*{8cm} \includegraphics{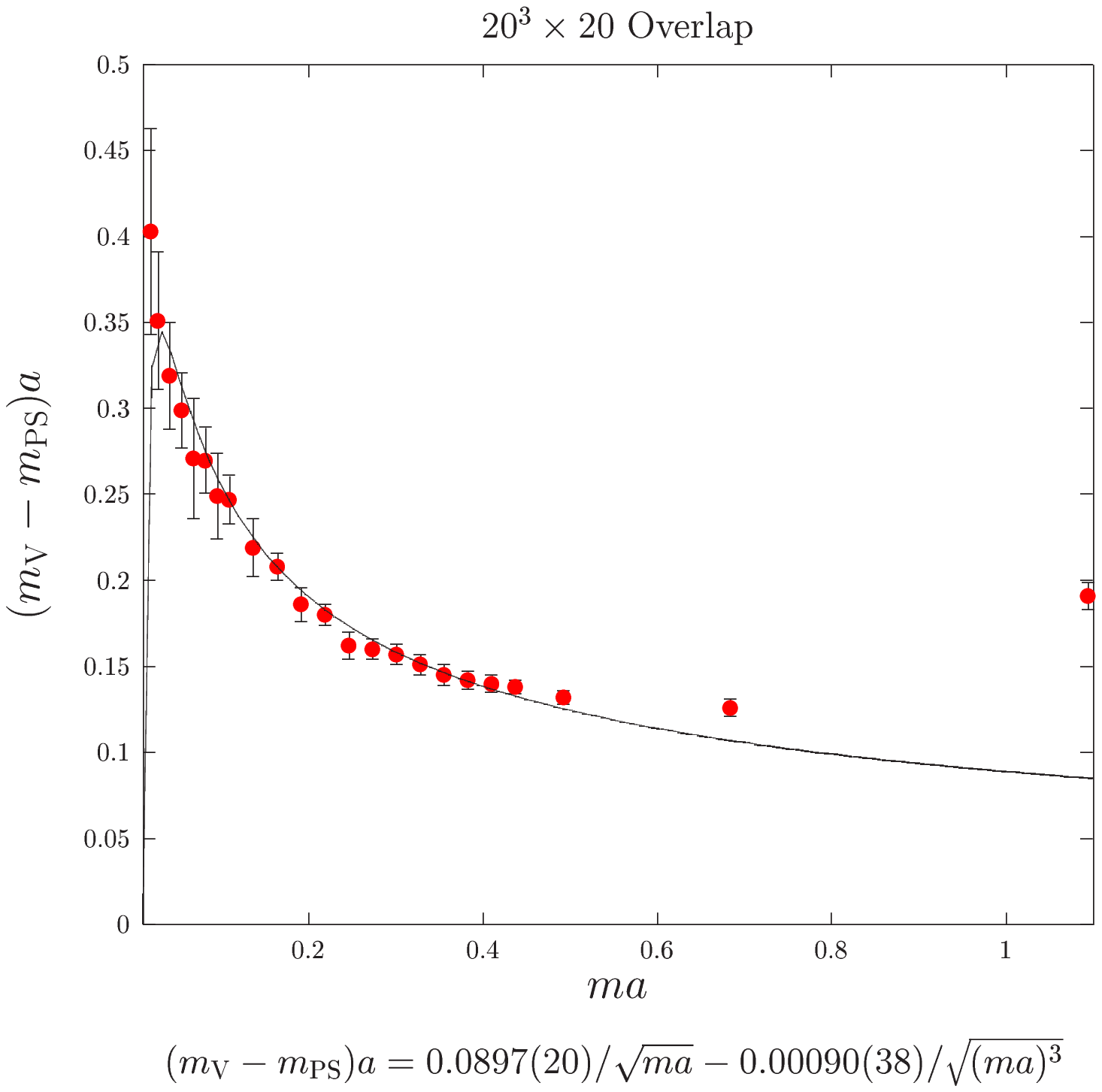} 
\caption{The hyperfine splitting on two lattices as a function of
$ma$. The upper panel is for the $16^3 \times 72$ lattice with $a
= 0.0561$ fm and the lower panel is for the $20^4$ lattice with
$a=0.133$ fm.} \label{hf_ma}
\end{figure}

 We first examine the $O(m^2a^2)$ and $O(m a^2)$ errors in the dispersion 
relation. It is suggested that dispersion relation is one of the places
where one can discern the $ma$ error~\cite{kla98,hmo02}. We
computed the pseudoscalar meson mass and energies at several
lattice momentum, i.e. $p_La = \sqrt{n}\, 2\pi/La$ with $n =
0,1,2,3$. The overlap quark propagators are calculated on the
$16^3 \times 28$ quenched lattice with 80 configurations generated
from Iwasaki guage action with $a = 2.00$ fm as determined from
$f_{\pi}$~\cite{cdd04}. Following Refs.~\cite{kla98,hmo02}, we fit
the energies to the dispersion relation
\begin{equation}
(E(p)a)^2 = c^2 (pa)^2 + (E(0)a)^2
\end{equation}
where $p = 2sin(p_La/2)$. The dispersion relation is so defined
such that the $ma$ error is reflected in the deviation of $c$ (the
effective speed of light) from unity.

   We see in Fig. \ref{disp_pi} that the effective
speed of light $c$ is consistent with unity all the way to
$ma \sim 0.4$. Since there is no $O(m a)$ error, we fit it with
the form quadratic in $a$, i.e. 
$c = c_0 + b\, (\Lambda_{QCD}a) ma
+ d\, m^2 a^2$ ($\Lambda_{QCD}a = 0.188$ for $a=0.2$ fm), and find that
$c_0=0.982(10),\, \mbox{b= 0.580(346)}$, and $d = -0.279(87)$ with
$\chi^2/N_{dof}= 0.1$ for the pseudoscalar meson case and
$c_0=1.044(43),\, b= 0.016(38)$, and $d = -0.41(36)$ with
$\chi^2/N_{dof}= 0.1$ for the vector meson case. 
Using these to gauge how large the $ma$
errors are, we see that the systematic error is less than $\sim 4\%$
for both the pseudoscalar and vector mesons up to $ma \sim 0.56$. This $m \, a$ is
$\sim 2.4$ times larger than that is admitted in the study of
improved Wilson action~\cite{hmo02} where it is found that the
$O(m^2 a^2)$ errors from the anisotropy of the dispersion relation for the
pseudoscalar and vector mesons are less than $\sim 5\%$ when $m_Q a_t < 0.23$.

\begin{figure}[h]
\vspace*{8cm} \includegraphics{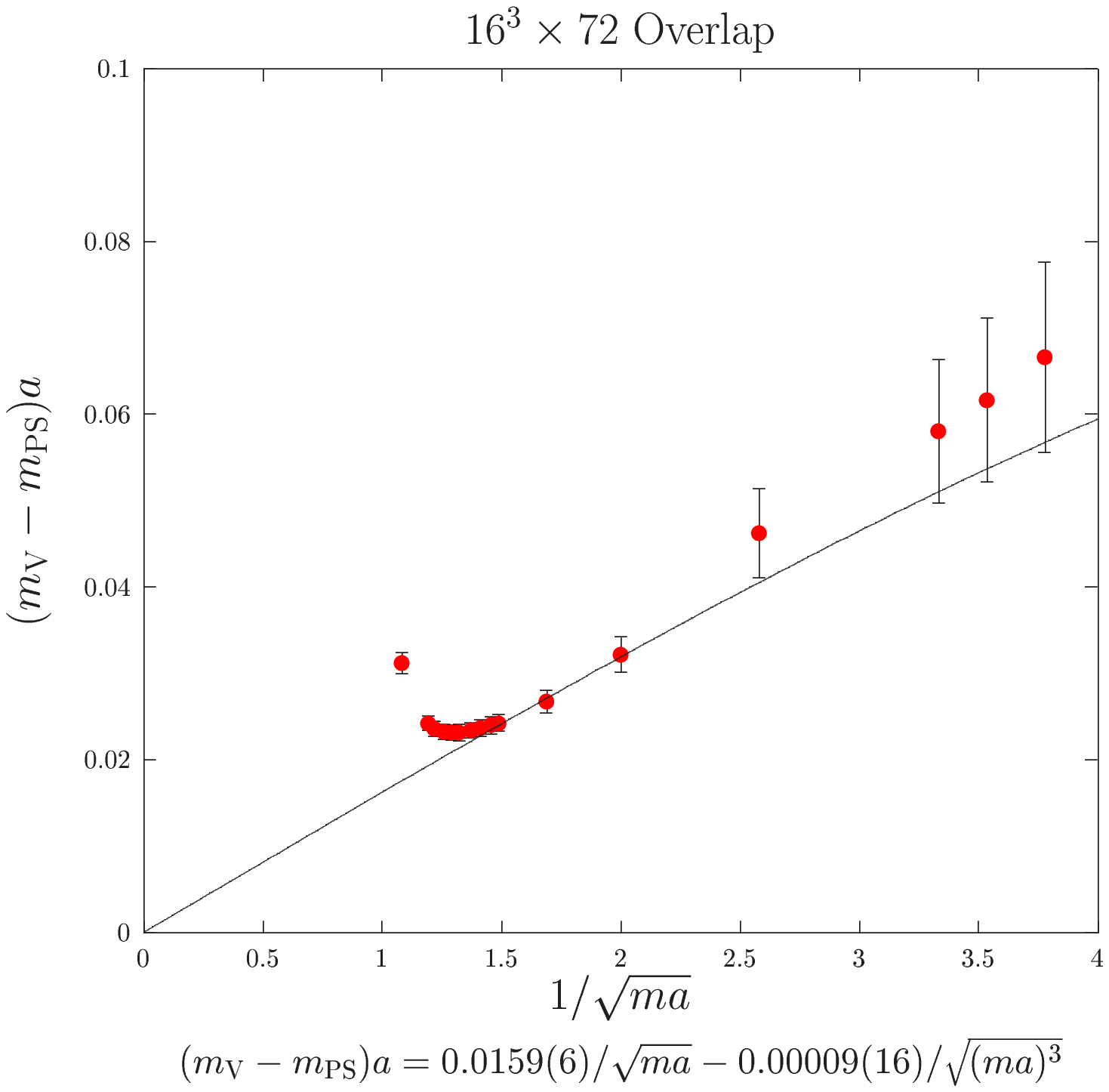} \vspace*{0.6cm}
\vspace*{8cm} \includegraphics{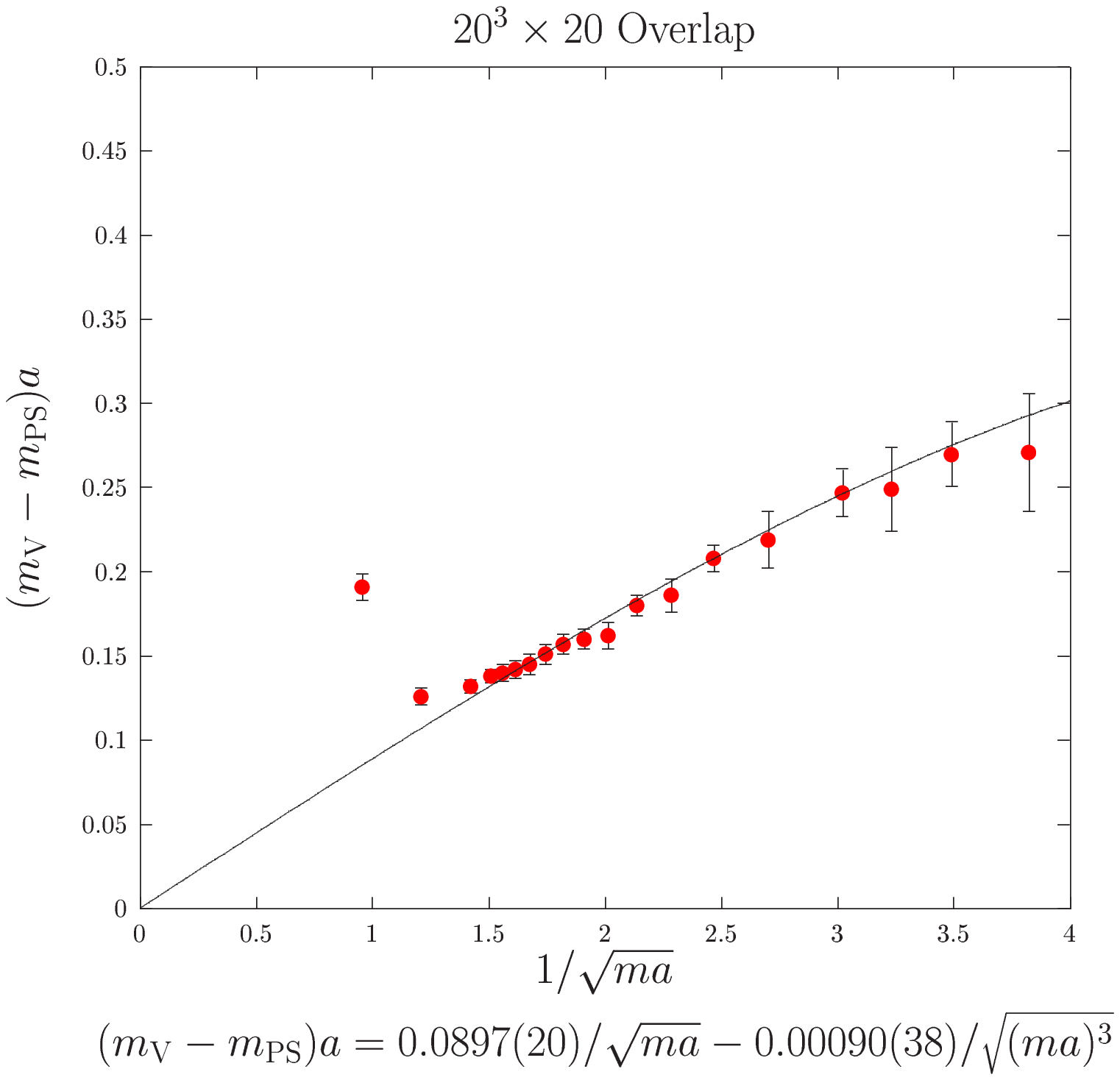} \vspace*{0.5cm}
\caption{The same as in Fig. 2 except as a function of $1/\sqrt{ma}$.}
\label{hf_sqrtma}
\end{figure}

   The other physical quantity we calculate is the hyperfine
splitting between the vector and pseudoscalar mesons as a function
of $ma$. We should first point out that this hyperfine splitting
is expected to go down with the square root of the quark mass for
heavy quarks~\cite{lw78}. This is so because the spin-spin part of
the one-gluon-exchange interaction which is expected to dominate
the short distance behavior between the heavy quarks in the
quarkonium has the form

\begin{equation}
V_{SS} \propto \frac{\alpha_s\lambda_1 \cdot \lambda_2}{m_1
m_2}\sigma_1 \cdot \sigma_2 \delta(\vec{r}_1-\vec{r}_2),
\end{equation}
which leads to a hyperfine splitting between the equal-mass vector
and pseudoscalar mesons in first order perturbation in $\alpha_s$
\begin{equation}
h.f.s \propto \frac{|\Psi(0)|^2}{m^2},
\end{equation}
where $\Psi(0)$ is the wavefunction of the quarkonium at the
origin. In view of the fact that the $2S-1S$ radial excitation and the
splitting between the averaged ${}^3P_{2}, {}^3P_{1}$ and ${}^3P_{0}$ and
the ${}^3S_1$ state (i.e. ${}^3P_{avg} - {}^3S_1$) of the vector mesons $J/\Psi$ and
$\Upsilon$ are almost the same, one deduce from the
non-relativistic potential model that the size of these mesons
scale like
\begin{equation}
r \propto \frac{1}{\sqrt{m}}
\end{equation}
in order to keep the excitation independent of the quark mass. Since
$\mid\Psi(0)\mid^2$ scales like $r^{-3}$, hence one obtains
\begin{equation}  \label{hfs}
h.f.s \propto \frac{1}{\sqrt{m}}.
\end{equation}

   We show in Fig.~\ref{hf_ma} the hyperfine splitting between
the vector meson and pseudoscalar meson for the $16^3 \times 72$
lattice with $a= 0.0561$ fm~\cite{tac06} and the $20^4$ lattice with $a= 0.133$
fm~\cite{ddh02} as a function of $ma$. We notice first that, despite of the fact
that the lattice spacings of these two lattices differ by a factor
of 2.37, their behaviors in $ma$ are very similar. Furthermore,
the hyperfine splittings in both cases do not approach zero at large 
quark mass as they should and this is obviously due to the $ma$ errors. 
To assess the errors, we plot in Fig.~\ref{hf_sqrtma} the hyperfine
splitting as a function of $1/\sqrt{ma}$. It is clear that there is
a broad range of $ma$ where the hyperfine splitting is largely
proportional to $\sqrt{ma}$ as in Eq.~(\ref{hfs}). But there are a
few outliers at large $ma$ which deviate substantially from the
$1/\sqrt{m}$\, behavior. These are due to the systematic $ma$ errors.
We fit the region which is largely linear in $1/\sqrt{ma}$ with a
form which also takes into account the $1/m$ correction, i. e.
\begin{equation}
h.f.s. = \frac{a}{\sqrt{ma}}(1 + \frac{b}{ma}).
\end{equation}
This form fits well in the range of $ma$ from 0.07 to 0.47 for the $16^3
\times 72$ lattice and from 0.1094 to 0.438 for the $20^4$
lattice. The fits are drawn as solid lines in Figs.~\ref{hf_ma}
and \ref{hf_sqrtma}. We see in both cases, the lattice results
start to deviate from the fits around $ma = 0.5$ and correspondingly
$1/\sqrt{ma}=1.4$. For $ma = 0.6$, the $m^2 a^2$ error is about 7\%. 
By the time $ma$ reaches 0.85, the $m^2a^2$ error is about 50\%. 

    By examining the $ma$ errors of the deviation from the
effective speed of light and the hyperfine splitting, we conclude
that it is prudent to use $ma$ smaller than 0.5 in the overlap 
fermion formalism in order to keep the systematic $O(ma^2)$ and 
$O(m^2a^2)$ errors to less than 3 to 4 \%. This study is done
with the Iwasaki gauge action. We have not explored
if and how this conclusion varies with different gauge actions.

This work is partially supported by DOE Grants DE-FG05-84ER40154
and DE-FG02-95ER40907.

\end{document}